\documentclass[%
 reprint,
 amsmath,amssymb,
 aps,
showpacs,
showkeys
]{revtex4-2}

\usepackage{graphicx}
\usepackage{dcolumn}
\usepackage{bm}
\usepackage{hyperref}

\def\GeV2 {$\mathrm{GeV} ^2$}
\def\ppbar{\mbox{$p \bar p$}}

\def\dsigel0 {$d\sigma_{el} /dt \vert _{t=0}$}

\begin{document}




\title{Results on Elastic Cross Sections in Proton--Proton Collisions at $\sqrt{s} = 510$~GeV with the STAR Detector at RHIC}

\author{
M.~I.~Abdulhamid$^{4}$,
B.~E.~Aboona$^{56}$,
J.~Adam$^{15}$,
L.~Adamczyk$^{2}$,
J.~R.~Adams$^{40}$,
I.~Aggarwal$^{42}$,
M.~M.~Aggarwal$^{42}$,
Z.~Ahammed$^{63}$,
E.~C.~Aschenauer$^{6}$,
S.~Aslam$^{27}$
% ...
}

\address{\rm{(STAR Collaboration)}}

% ...
\address{$^{2}$AGH University of Krakow, FPACS, Cracow 30-059, Poland}
% ...
\address{$^{4}$American University in Cairo, New Cairo 11835, Egypt}
% ...
\address{$^{6}$Brookhaven National Laboratory, Upton, New York 11973}
% ...
\address{$^{15}$Czech Technical University in Prague, FNSPE, Prague 115 19, Czech Republic}
% ...
\address{$^{27}$Indian Institute Technology, Patna, Bihar 801106, India}
% ...
\address{$^{40}$The Ohio State University, Columbus, Ohio 43210}
% ...
\address{$^{42}$Panjab University, Chandigarh 160014, India}
% ...
\address{$^{56}$Texas A\&M University, College Station, Texas 77843}
% ...
\address{$^{63}$Variable Energy Cyclotron Centre, Kolkata 700064, India}
% ...

\begin{abstract}
We report results on an elastic cross section measurement in proton--proton collisions at a center-of-mass energy $\sqrt{s}=510$~GeV, obtained with the Roman Pot setup of the STAR experiment at the Relativistic Heavy Ion Collider (RHIC).
The elastic differential cross section is measured in the four-momentum transfer squared range $0.23 \leq -t \leq 0.67$~GeV$^2$.  This is the only measurement of the proton-proton elastic cross section in this $t$ range for collision energies above the Intersecting Storage Rings (ISR) and below the Large Hadron Collider (LHC) colliders.
We find that a constant slope $B$ does not fit the data in the aforementioned $t$ range, and we obtain a much better fit using a second-order polynomial for $B(t)$. This is the first measurement below the LHC energies for which the non-constant behavior $B(t)$ is observed. 
The $t$ dependence of $B$ is also determined using six subintervals of $t$ in the STAR measured $t$ range, and is in good agreement with the phenomenological models. The measured elastic differential cross section $\mathrm{d}\sigma/\mathrm{dt}$ agrees well with the results obtained at $\sqrt{s} = 540$~GeV for proton--antiproton collisions by the UA4 experiment. 
We also determine that the integrated elastic cross section within the STAR $t$-range is $\sigma^\mathrm{fid}_\mathrm{el} = 462.1 \pm 0.9 (\mathrm{stat.}) \pm 1.1 (\mathrm {syst.}) \pm 11.6 (\mathrm {scale})$~$\mu\mathrm{b}$.
\end{abstract}
\maketitle

\section{Introduction}\label{sec:Intro}

Most of the proton--proton ($pp$) elastic scattering cross-section measurements are in the four-momentum transfer squared $t$ range where perturbative QCD (pQCD) cannot be applied. Here, $t = (p_\mathrm{in} - p_\mathrm{out} )^2$, where $p_\mathrm{in}$, $p_\mathrm{out}$ represent the four-momenta of the incoming and outgoing proton, respectively. 
Unlike the case for pQCD, where the QCD Lagrangian is used to calculate the scattering amplitudes, the calculations in the low $t$ range are done in the Regge framework~\cite{Regge:1959mz,ReggeIntro,DiffractionBarone,ReggeTrajectory}, 
where the amplitudes are evaluated in the framework of scattering matrix (S-Matrix) theory.
Those scattering amplitudes ${\cal A}(s,t)$ depend on the square of center-of-mass energy $s$, and $t$. Regge theory provides rigorous constraints on the properties of the scattering amplitudes~${\cal A}(s,t)$.

In the $t$ range of this measurement, $0.23 \leq -t \leq 0.67$~GeV$^2$, the elastic cross section $d\sigma/dt$ is described by the hadronic term of the scattering amplitude ${\cal A}(s,t)$ with an exponential dependence on $t$: $d\sigma/dt = |{\cal A}(s,t)|^2 = A \cdot e^{-B(t) \,|t|}$. Although the theory allows for the exponential slope $B$ to depend on $t$, the data show that at a given $\sqrt{s}$ the slope is approximately constant for small $|t|$ but changes at large $|t|$. For example, there is a well-known change in slope at $|t| \approx 0.13$~GeV$^2$ as discussed in \cite{Matthiae:1994uw}.

At $\sqrt{s} \gtrsim 10$~GeV energies, depending on $\sqrt{s}$, the elastic scattering contributes 18 -- 28\% to the total cross section. Hence, it is important to measure it at every available $\sqrt{s}$. Each new data set provides additional information, which is then used in the tuning of phenomenological models of elastic scattering. If measured at low enough $t$, the elastic cross section allows a determination of the total cross section. For these reasons, elastic scattering has typically been measured at all particle accelerator facilities. 

This paper reports the results on $pp$ elastic scattering at $\sqrt{s}=510$~GeV, which is below those most recently measured at the LHC with center-of-mass energies $2.76 \le \sqrt{s} \le 13$~TeV \cite{TOTEM:2017asr,TOTEM:2013vij,TOTEM:2016lxj,TOTEM:2018psk,ATLAS:2014vxr,ATLAS:2016ikn,TOTEM:2015oop,ATLAS:2022mgx}.
It is above the $\sqrt s$ range of the Intersecting Storage Rings (ISR) measurements carried out about 50 years ago at $\sqrt{s}=62.4$~GeV \cite{CERN-Pisa-Rome-StonyBrook:1976mtu,Breakstone:1985pe}
and a recent STAR measurement~\cite{STAR:2020phn}
of $pp$ elastic scattering in a lower $|t|$ range at $\sqrt s = 200$~GeV. 
This is the first measurement below the LHC energies for which a non-constant behavior $B(t)$ is observed. It is also in a different $t$ range than that reported by TOTEM and ATLAS collaborations \cite{TOTEM:2015oop,ATLAS:2022mgx} at the LHC.
The $\ppbar$ elastic scattering was measured at the ISR, the $S\!\ppbar S$ collider at $\sqrt s  = 540$ and $630$~GeV \cite{UA4:1983mlb,UA4:1985oqn,UA4:1986cgb} 
and at the Tevatron at 1.8 TeV and 1.96~TeV \cite{E710:1991bcl,CDF:1993qdf,D0:2012erd}.
In particular, the $S\!\ppbar S$ UA4 experiment at $\sqrt s  = 540$~GeV~\cite{UA4:1983mlb} found a constant $B$-slope of $13.7 \pm 0.3$~GeV$^{-2}$ in $t$-range $0.21 \leq -t \leq 0.50$~GeV$^2$, similar to STAR. 

\section{The Experiment}\label{sec:ExpSetup}

\begin{figure}[!b]
\includegraphics[width=1.\columnwidth]{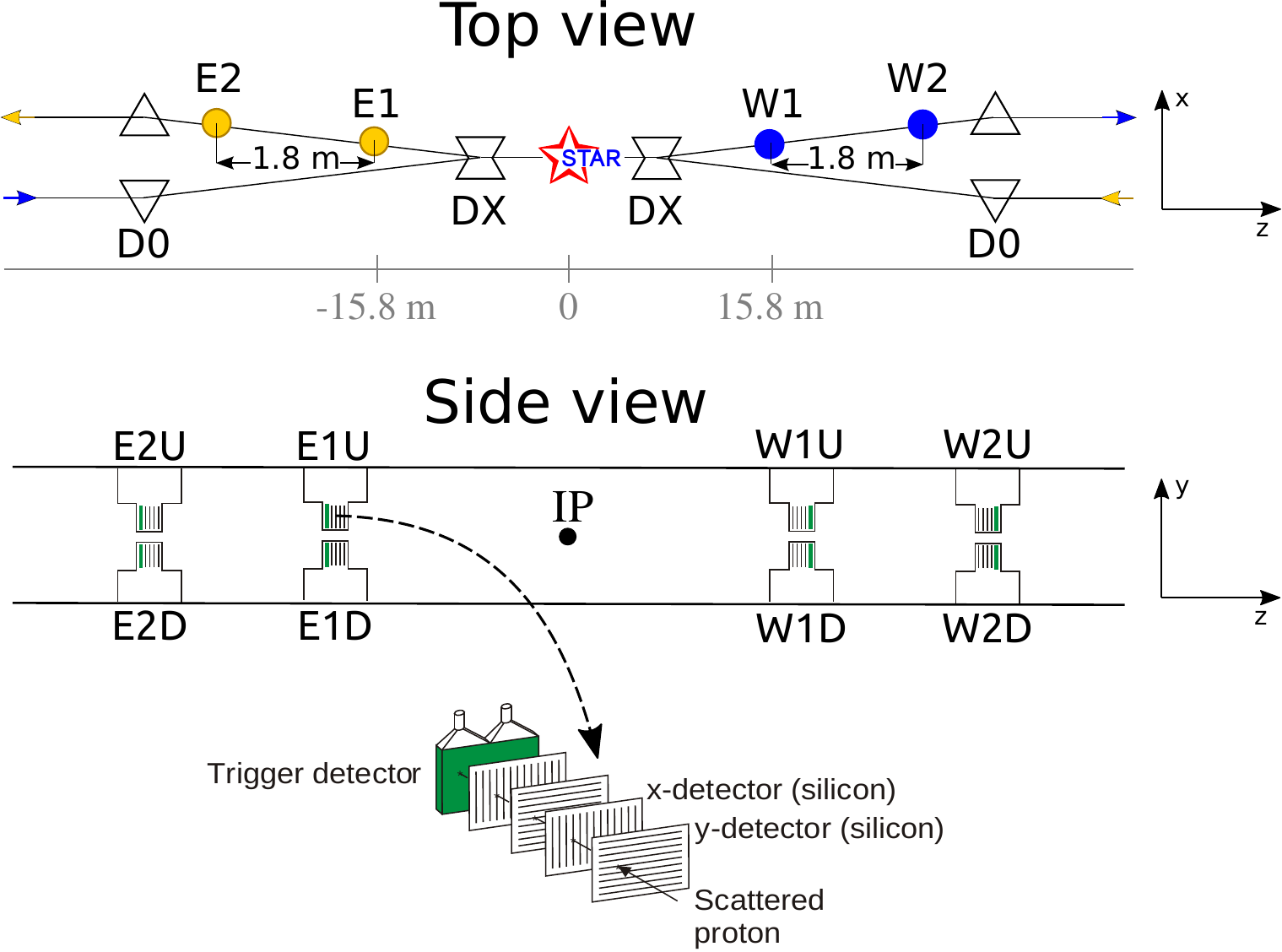}
\caption{The layout of the RP setup at STAR (not to scale) for measuring forward protons. Top $(x,z)$ and side $(y,z)$ views are shown.
Two sets of RPs, labeled (W1, W2) and (E1, E2) were installed between the DX and D0 magnets, at 15.8~m and 17.6~m, on either side of the IP. The detector package has transverse size $5\times 8\; \mathrm{cm}^2$ and a depth 3.5~cm. The Si sensor is $400\;\mu\mathrm{m}$ thick, while the trigger scintillator is 5 mm thick. The strips in the Si detectors are $\approx 100\; \mu$m wide. Two dipole magnets, DX and D0, which bend the beams into and out of the IP, are also shown.}
\label{fig:Figure1-510GeV}
\end{figure}

The results presented here are obtained with the setup described in \cite{STAR:2020phn}, 
whose main features are described below. For these measurements at $\sqrt{s}=510$~GeV, the STAR experiment~\cite{STAR:2002eio} 
was upgraded with the Roman Pot (RP) system used previously by the PP2PP experiment~\cite{Bultmann:2004ke}. 
The location of the RPs, top and side views, and the coordinate system are shown schematically in Fig.~\ref{fig:Figure1-510GeV}. Each RP station contains four Silicon (Si) strip detectors and a trigger scintillation counter. 
The elastic scattering is determined in the  STAR coordinate system, where the $z$-axis is in the direction of the clockwise-going RHIC beam, the $y$-axis is pointing up and the $x$-axis completes the right-handed coordinate system whose origin is at the interaction point (IP).

The DX magnet, the RHIC-lattice dipole magnet closest to the IP, and the detectors in the two sets of RPs enable the measurement of the momentum vector of the scattered protons at the detection point. Using that information the scattering angle at the IP is determined.
Because of the symmetry of the RHIC rings, the fields in the DX magnets on both sides of the IP are identical at the $10^{-3}$ level. Therefore, the bending angles of the magnets are also the same.

The data for the results reported here were acquired in the RHIC 2017 run during the period with a special accelerator optics with $\beta^*\approx 8$~m, (where $\beta^*$ is the $\beta$-function value at the collision point), which resulted in a beam angular divergence of $\approx 30$ $\mu$rad, which is smaller than that during the standard running conditions. The emittance in both $x$ and $y$ were the same and constant within 10$\,-$15\% during the data taking, as determined by beam monitors.
Luminosity monitors were calibrated using Van der Meer scans~\cite{vanderMeer:1968zz}.
The range of instantaneous luminosity was $6\: \mbox{to}\: 13\cdot 10^{30}\:\mbox{cm}^{-2}\mbox{s}^{-1}$. The associated systematic uncertainty on the luminosity measurement~is~2.2\%.

The RPs were moved as close to the beam as possible; the closest position of the first readout strip was about 20 mm, which corresponds to a minimum $|t|$ of about 0.16 $\mbox{GeV}^2$.
The aperture of the DX magnet and the following beam pipe structure determined a maximum achievable value of $|t| \approx 1.1\, \mbox{GeV}^2$, corresponding to a scattering angle of $\theta \approx 4~\mbox{mrad}$. In this paper, we analyze the elastic scattering in the region of uniform geometrical acceptance in the range $0.23 \leq -t \leq 0.67$~GeV$^2$. This allows us to minimize the background due to beam halo and scattering on the apertures.

There are about $26\cdot10^6$ triggered events for the integrated luminosity of $304\; \mbox{nb}^{-1}$. They satisfy the elastic scattering trigger condition: 
\begin{equation}
 \mathbf{( EU \land WD ) \lor ( ED \land WU )},
 \label{eq:TriggerCond}
\end{equation}
\noindent where EU denotes a valid PMT signal in at least one of the PMTs of the EU1 or EU2 trigger counters. 
Similarly, ED, WU and WD denote valid PMT signals in the other trigger counters, as shown in Fig.~\ref{fig:Figure1-510GeV}. 
\section{Clustering, Track Reconstruction and Alignment}\label{sec:AlignmTrack}
Track reconstruction in the Si detectors is a multi-step process. Initially, clustering is used to determine the position of the proton trajectory in a Si plane. Then, the reconstruction of a point (PT) in a RP is performed. Finally, the scattering angles are reconstructed and the $t$ value is determined.
\subsection{Clustering}\label{sec:Clustering}
First, to make sure that the deposited energy in a Si strip is above the noise, the energy measured in that strip is required to be larger than $5\sigma_\mathrm{ped}$, where $\sigma_\mathrm{ped}$ is the average pedestal width of the 126 channels in one readout SVXIIE chip \cite{Zimmerman:1995bq}.

Second, a clustering procedure for each Si plane is performed following Ref.~\cite{STAR:2020phn}.
However, in this analysis, there is a minimum energy cut, which depends on the cluster length, i.e., the number of consecutive strips in the cluster. Clusters longer than 5 strips are excluded. These cluster energy and cluster size cuts, determined in a data-driven way, are used to suppress background. 
The signal-to-noise ratio is about $20 :\! 1$, as measured by the Most Probable Value of the Landau $dE/dx$ distribution for a cluster size of one strip and is found to be larger for larger clusters.

Third, matching of the clusters between planes measuring the same coordinate is performed to reconstruct a PT's $(x,y)$ coordinates in a given RP. The clusters are considered matched if the distance between them is less than 300 $\mu$m. In case a cluster is found in only one of the planes for a given coordinate, that coordinate is used only if there are matched clusters in the other coordinate. These PTs are used to reconstruct the scattering angles.
\subsection{Track and Scattering Angle}\label{sec:TrackAngle}
Two points reconstructed on the same side of the IP, one in each RP, define a track. The scattering angles $(\theta_x, \theta_y)$, in the $(x,z)$ and in the $(y,z)$, plane of that track are calculated using those two PTs:
\begin{equation}
 \theta_x = \frac{X_\mathrm{RP2} - X_\mathrm{RP1}}{Z_\mathrm{RP2} - Z_\mathrm{RP1}},   \mathrm{\hspace{1cm}} \theta_y = \frac{Y_\mathrm{RP2} - Y_\mathrm{RP1}}{Z_\mathrm{RP2} - Z_\mathrm{RP1}},
 \label{eq:ThetaXYReco}
\end{equation}
where RP1 and RP2 denote near and far RP stations with respect to the IP. The coordinates $(X_\mathrm{RP}, Y_\mathrm{RP})$ are with respect to the nominal beam trajectory. The $Z_\mathrm{RP}$ is the $z$-position of the RP with respect to the IP.

About $70\%$ of the events had one and only one PT per RP on the upper (lower) East or West side of the IP.
Alignment is performed for each run in the analysis using the procedure described in Ref.~\cite{STAR:2020phn}.
The resulting run-by-run corrections to the positions of the strips are applied before the reconstruction of the scattering angles. As such, the alignment offsets are obtained in the system of coordinates where the two protons are elastically scattered, a collinear elastic scattering geometry.

\section{Data Analysis}\label{sec:DataAnal}

In this section, we describe the flow of the data analysis. The scattering angles $\theta_x$ and $\theta_y$ are calculated from the points reconstructed in the Si detectors, as described above. Then cuts are applied to select elastic scattering events. 
\subsection{Analysis Selection Criteria}\label{sec:AnalCuts}
The various selection criteria for choosing elastic events are described below in the order as they are applied in the analysis:

{\bf Elastic event topology (ET)}: Only events with a combination of reconstructed points in the RPs consistent with elastic scattering are accepted. Namely, the combinations with the lower East detector in coincidence with the upper West detector, arm EDWU, or the upper East detector in coincidence with the lower West detector, arm EUWD, satisfy the elastic event topology due to momentum conservation. 6.33 M events remained after this cut.

{\bf Four Roman Pot (4RP) event data sample}: Only events with at least one reconstructed point per RP on the East and on the West are kept. 1.95 M events remained after this cut.

{\bf Four PT (4PT) events}: 4RP events with one and only one PT per RP and no reconstructed points in the Si in the other arm. Using 4PT events the scattering angles ($\theta_x, \theta_y)$ on each side of the IP are calculated as indicated in Eq.~\ref{eq:ThetaXYReco}. 1.63 M events remained after this cut.

\begin{figure}[!t]
\begin{center}
\includegraphics[width=1.\columnwidth]{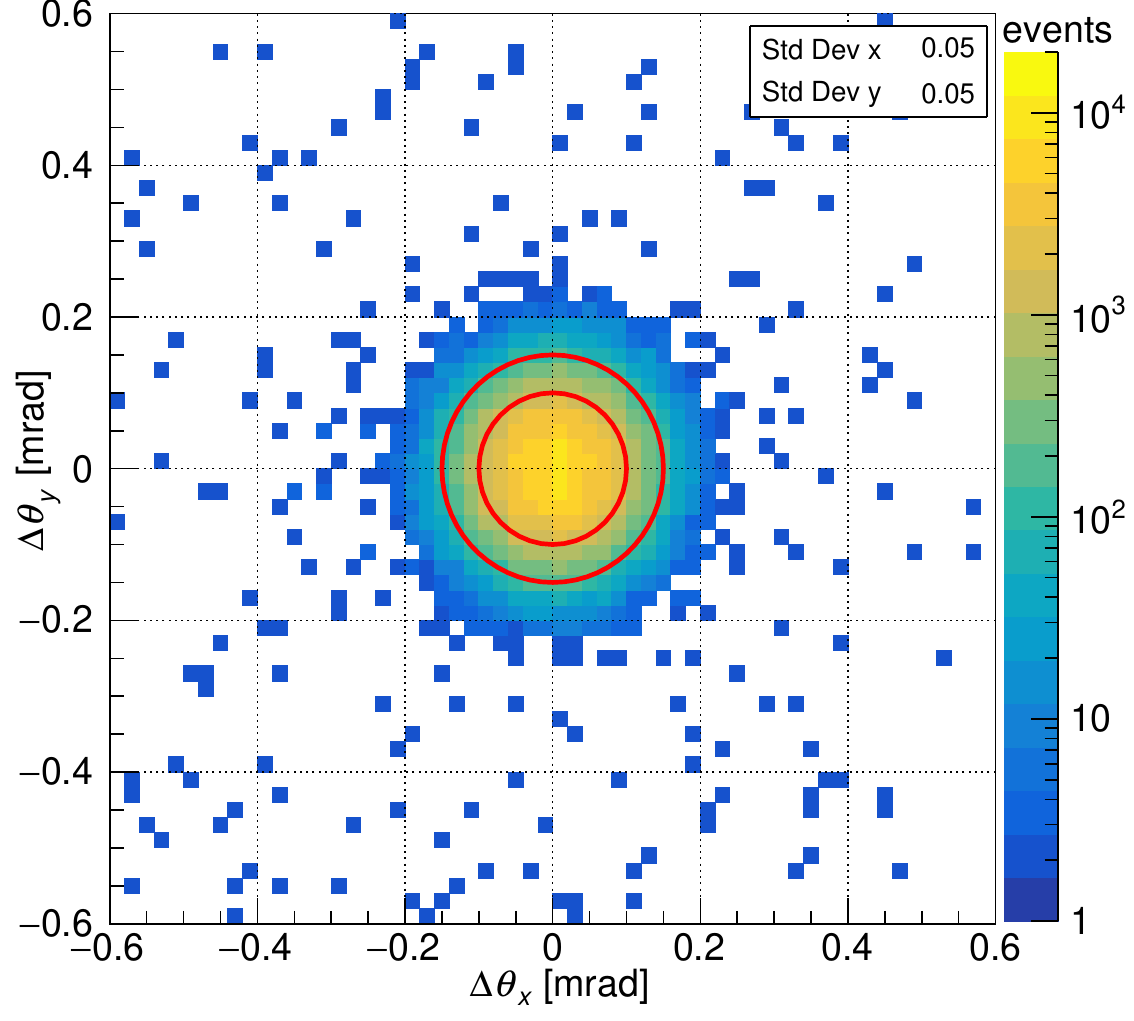}
\caption{The collinearity of the data sample for accepted ET events is shown. It is defined as the differences $\Delta\theta_x$ and $\Delta\theta_y$ between the scattering angles $\theta_x$, $\theta_y$ reconstructed on the East and West side of the IP. Contours of $2\sigma_{\Delta\theta}$ and $3\sigma_{\Delta\theta}$, where $\sigma_{\Delta\theta} \approx \sigma_{\Delta\theta_x} \approx \sigma_{\Delta\theta_y} \approx 50 ~\mu\mbox{rad}$, are shown as red circles.}
\label{fig:Figure2-510GeV}
\end{center}
\end{figure}

{\bf Collinear (COL) events}: The $\theta_{W}$ and $\theta_{E}$ are the reconstructed polar scattering angles on the West and East sides of the IP, respectively. Because of momentum conservation, collinearity in  $\theta_{W}$, $\theta_{E}$ is required. Hence, $\Delta\theta = \theta_{W} - \theta_{E}$ is expected to be zero. Consequently, we select the events for which $|\Delta\theta| \le 3\sigma_{\Delta\theta}$, where $\sigma_{\Delta\theta} = 50 $ $\mu\mbox{rad}$ is the Gaussian width of the collinearity distribution, consistent with the beam angular divergence. 
The collinearity condition also requires that the distance between the two projected tracks in $x$ and $y$ at $z = 0$ be within a radius of $3\sigma$ of the Gaussian width of their distributions.
In Fig.~\ref{fig:Figure2-510GeV}, we show the collinearity distribution $\Delta\theta_y$ vs $\Delta\theta_x$, where $\Delta\theta_x=\theta_x^W - \theta_x^E$ and $\Delta\theta_y=\theta_y^W - \theta_y^E$.
Here, the $\theta_x^W, \theta_x^E, \theta_y^W, \theta_y^E$ are scattering angles reconstructed on the West and East sides of the IP, using the measured coordinates at the RP and after the fiducial volume cut. A clear peak of elastic events is seen. 
The contours of $2\sigma_{\Delta\theta}$ and $3\sigma_{\Delta\theta}$ are also shown. 1.19 M events remained after this cut.

\begin{figure}[!t]
\begin{center}
\includegraphics[width=1.\columnwidth]{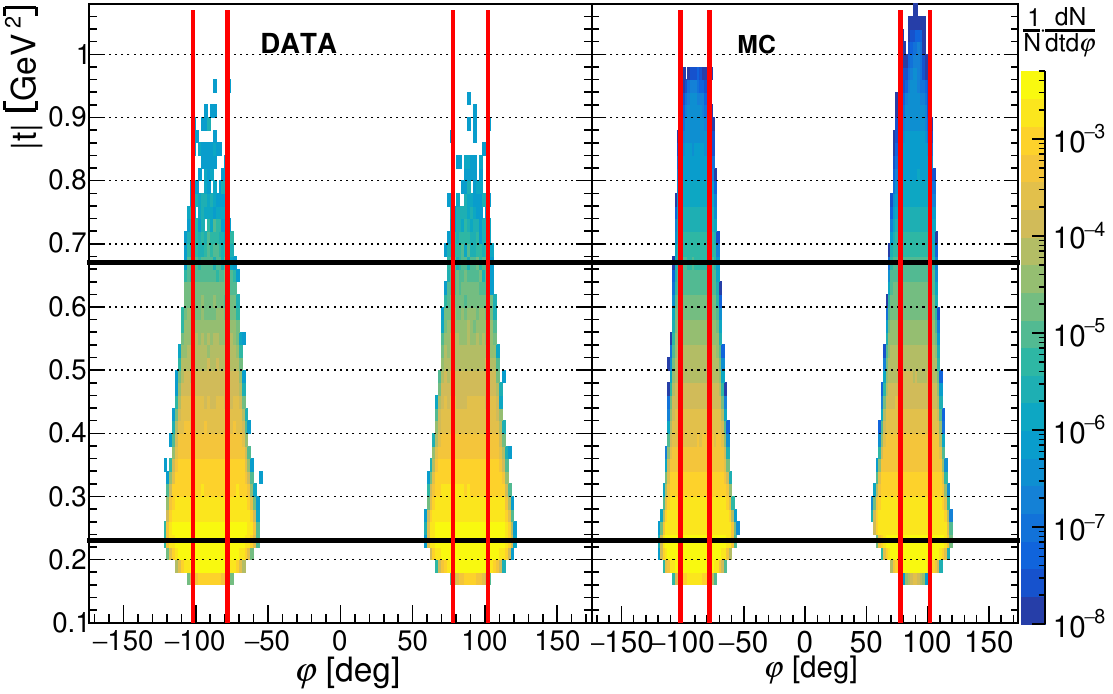}
\caption{Four-momentum transfer $|t|$ vs $\phi$ distributions for data and MC for 4PT collinear events.
 The two elastic combinations of tracks between East and West, EUWD and EDWU, are shown. The GEO cuts in $t$ and $\phi$ are shown as black and red lines, respectively.}
\label{fig:Figure3-510GeV}
\end{center}
\end{figure}

{\bf Fiducial volume (GEO) cut}: After the elastic event candidates are chosen based on collinearity, one more set of cuts in a fiducial volume $(\phi, |t|)$, where $\phi$ is the azimuthal angle of the proton scattering, is needed to remove the remaining background. To stay away from the beam halo, a minimum $|t|$ corresponding to $12\sigma$ of the beam size is required, well outside of the beam envelope. Hence, a coincidence arising from the beam halo from the two beams is not expected.
To stay away from the apertures, additional cuts on the maximum $|t|$ and on the $\phi$-range are also required. 
The chosen $\phi$, $|t|$ ranges are $78^\circ \leq |\phi| \leq 102^\circ$ and $0.23 \leq -t \leq 0.67$~GeV$^2$, respectively. 
The fiducial cuts are shown in Fig.~\ref{fig:Figure3-510GeV}.
These cuts are chosen based on the simulation, which is described in Sec.~\ref{sec:GeoTrackEff}. 0.35 M events remained after this cut.

\subsection{The $t$ Reconstruction}
The scattering angles $\theta_x$ and $\theta_y$ are determined by fitting a straight line using 4PT events and $\chi^2$ minimization.
Given the beam momentum $p$ and small  scattering angles $\theta_x$ and $\theta_y$, the $t$-value is calculated using:
\begin{equation}
t = (p_\mathrm{in} - p_\mathrm{out} )^2 \approx - p^2\theta^2 = - p^2\cdot ( \theta_x^2 + \theta_y^2).
\label{eq:t-def}
\end{equation}
The resolution in $t$ is dominated by the beam angular divergence, which is about $30 ~\mu\mbox{rad}$ for both $\theta_x$ and $\theta_y$,  as given by the beam emittance and by the $\beta$-function value at the collision point ($\beta^*$). The detector spacial resolution is a small fraction of the $t$ resolution. The measured standard deviations of the angular distributions of $(\Delta\theta_x, \Delta\theta_y)$ are $\thicksim 50 ~\mu\mbox{rad}$, as shown in Fig.~\ref{fig:Figure2-510GeV}. They are consistent with the estimate of the beam angular divergence and position reconstruction resolution in the Si detectors. The beam momentum resolution was at the $10^{-3}$ level, hence it was a small fraction of the $t$ resolution.
\section{Efficiency Corrections}\label{sec:Efficiency}
The efficiency correction has two terms: 1) efficiency which accounts for the limited geometrical acceptance and point reconstruction efficiency in a RP; 2) the trigger efficiency.
The former has two components: a MC component, which accounts mainly for geometrical acceptance, and a data-driven $t$-dependent point reconstruction efficiency within the geometrical acceptance. The trigger efficiency is obtained from the data using Zero Bias (ZB) triggers, which are events triggered on beam crossings only. 
%

We introduce a correction function $C(t)$, which relates the number of reconstructed elastic events $N_\mathrm{rec}^\mathrm{DATA}(t)$ obtained from data to the number of events produced at the vertex 
$N_\mathrm{cor}^\mathrm{DATA}(t)$:
\begin{equation}
\label{eq:CorrectionForm}
N_\mathrm{cor}^\mathrm{DATA}(t) =  N_\mathrm{rec}^\mathrm{DATA}(t) \cdot C(t).
\end{equation}
\noindent The corrections from which $C(t)$ is obtained are discussed in the following sections.
\subsection{Geometrical Acceptance and Track Reconstruction Efficiency}\label{sec:GeoTrackEff}
To determine the $t$-dependent geometrical acceptance of the RP detector system, a GEANT4-based~\cite{GEANT4:2002zbu} 
simulation is used. The simulation includes a detailed description of the DX magnet including all limiting apertures, the RP details, and the Si readout behavior. The latter includes known hardware problems, such as the two non-working (out of 80) SVX readout chips and one non-working (out of 32) Si plane. The two non-working SVXs were mostly outside of the geometrical acceptance. The energy deposited by final state particles in the Si detectors is digitized and added to the electronic noise obtained from the pedestal runs.
To reproduce the impact of background, the MC-generated events are embedded in the ZB data sample. This is done by combining the list of clusters from the ZB events with the list of simulated clusters. The overlaying clusters are merged and their positions are recalculated. After the embedding, a standard PT reconstruction, including cluster matching, is done the same way as in the real data.
The $pp$ elastic scattering is generated using uniform distributions in $t$ and $\phi$ ranges of $0.1 \leq |t| \leq 1.5 ~\mbox{GeV}^2$ and $-\pi\leq \phi < \pi$, respectively. As a result, the geometrical acceptance of the detector is obtained as the main contribution to the efficiency correction function defined as:
\begin{equation}
\varepsilon_\mathrm{MC} (t) = \frac{N^\mathrm{MC}_\mathrm{rec}(t) }{N^\mathrm{MC}_\mathrm{gen}(t)} \ or \  C_\mathrm{MC}(t) = \frac{1}{\varepsilon_\mathrm{MC} (t)},
 \label{eq:CorrFactor}
\end{equation}
where $N^\mathrm{MC}_\mathrm{gen}(t)$ and $N^\mathrm{MC}_\mathrm{rec}(t)$ are the true and reconstructed distributions obtained as functions of generated and reconstructed $t$, respectively.

 That purely geometrical acceptance factor, based on an angular acceptance $\Delta\phi = \pm12^\circ$, is in first order $C_\mathrm{MC} =360^\circ/24^\circ = 15$. Furthermore, to account for the fact that the MC events are generated with flat distributions, the data are reweighted event by event using the FMO model~\cite{Martynov:2018sga}.
 The systematic effect of the reweighting procedure is estimated in Sec.~\ref{sec:reweighting}.

The efficiency of point reconstruction for each RP is estimated using the data sub-sample containing only events with one reconstructed point in each of the three RPs, not including the RP under the test.
A track is reconstructed using those three points. The track has to pass the GEO filters to ensure that it crosses the geometrical acceptance of each RP, and then is projected to the RP under test. 
If the distance $D$ between the projected position of the track and the reconstructed point in that RP is less than $1.5$~mm and the reconstructed 4PT event satisfies the criteria for an elastic event, that RP is considered efficient and the count is added to the $N_\mathrm{pass}$ sample. If the event does not satisfy those criteria, the count is added to the $N_\mathrm{fail}$ sample. The PT reconstruction efficiency is then obtained as the ratio of the number of tracks crossing a given RP with the reconstructed PT found in this RP to the number of all tracks crossing the RP, and measured as a function of reconstructed $t$:
\begin{equation}\label{eq:effiRP}
 \varepsilon^\mathrm{PT}_\mathrm{RP}(t) = \frac{N_\mathrm{pass}(t)}{N_\mathrm{pass}(t)+N_\mathrm{fail}(t)}.
\end{equation}

More than 99\% of the events projected into a RP satisfy the $D \le 1.5$~mm cut. Hence, this cut does not affect the event reconstruction efficiency. This criterion is only used to select possible 4PT event candidates, for which collinearity (COL) and geometrical acceptance (GEO) criteria for an elastic event are checked.

The four $\varepsilon^\mathrm{PT}_\mathrm{RP}(t)$ efficiencies are not independent. In addition to the four separate cases where events are lost due to either no point or more than one point in the tested RP, there is a common case for all four RP's. This occurs when a 4PT event made of single reconstructed points does not pass the standard elastic selection imposed on 4PT events, resulting in a  $\varepsilon^\mathrm{COL}_\mathrm{RP}(t)$ for each RP. Since we find this value to be constant for all RPs, 
that common factor $\varepsilon_\mathrm{COL}(t) \approx 0.98$ is used as the overall correction factor.

The efficiency for each arm $\varepsilon_\mathrm{arm}(t)$, EUWD or EDWU combinations of the RPs, is then obtained as the product of the above five independent efficiencies in that arm $\varepsilon_\mathrm{arm}(t) = \varepsilon_\mathrm{4PT}(t)\cdot \varepsilon_\mathrm{COL}(t)$, where $\varepsilon_\mathrm{4PT}(t)$ is a product of the four efficiencies of finding a point (PT) in each RP before a collinearity cut is applied.
The same procedure as for the data is also used for the MC-embedded sample. The difference between the two is at the level of a few percent per RP, within its geometrical acceptance, and is interpreted as the effect of unknown inefficiencies present in the data and not included in the MC simulation (e.g., caused mainly by either no point or more than one point reconstructed in a single RP). It should be noted that no point in most cases means that too large a cluster was observed which is then not classified as a reconstructed point.
To account for that, a correction function $C_\mathrm{T}(t)$ to tune $\varepsilon_\mathrm{MC}$ is used:
\begin{equation}\label{eq:4PTCF}
  C_\mathrm{T}(t) = \frac{\varepsilon_\mathrm{arm}^\mathrm{MC-EMBD}(t)}{\varepsilon_\mathrm{arm}^\mathrm{DATA}(t)},
\end{equation}
where $\varepsilon_\mathrm{arm}^\mathrm{MC-EMBD}$ and $\varepsilon_\mathrm{arm}^\mathrm{DATA}$ are the elastic event reconstruction efficiencies obtained from the MC embedded sample and the data, respectively.

The systematic uncertainty on $\varepsilon^\mathrm{PT}_\mathrm{RP}(t)$ is estimated by varying the collinearity cut for 3PT events used to calculate the RP efficiency from the nominal $3\sigma$ by $\pm 1\sigma$.
\subsection{Trigger Efficiency}\label{sec:TrigEff}
The elastic event data stream contains only events triggered by the coincidence of valid PMT signals consistent with the trigger condition in Eq.~\ref{eq:TriggerCond}. The trigger efficiency is defined as the ratio of the number of events reconstructed with the silicon detectors satisfying the pattern in Eq.~\ref{eq:TriggerCond} and confirmed by the PMT trigger ($N_\mathrm{trig}^\mathrm{rec}$), over the number of all reconstructed events ($N^\mathrm{rec}_\mathrm{all}$) fulfilling the trigger condition:
\begin{equation}
\varepsilon_\mathrm{trig}(t) = \frac{N^\mathrm{rec}_\mathrm{trig}(t) }{N^\mathrm{rec}_\mathrm{all}(t)}.
\label{eq:trigefi}
\end{equation}
\noindent The trigger efficiency is calculated using the ZB data sample by comparing the trigger bit with the combination of PMT signals in a given event. 
A constant value $\varepsilon_\mathrm{trig}=0.986^{\scriptstyle +0.008}_{\scriptstyle -0.015}$ is used, as obtained by integration over the acceptance of this measurement.  
The quoted statistical uncertainty of the trigger efficiency is treated as an independent source of the overall normalization uncertainty and is added in quadrature to the other normalization
uncertainties.
\subsection{The Correction Function}\label{sec:FullCorrFun}
\begin{figure}[htp]
\centering
 \includegraphics[width=1.\columnwidth]{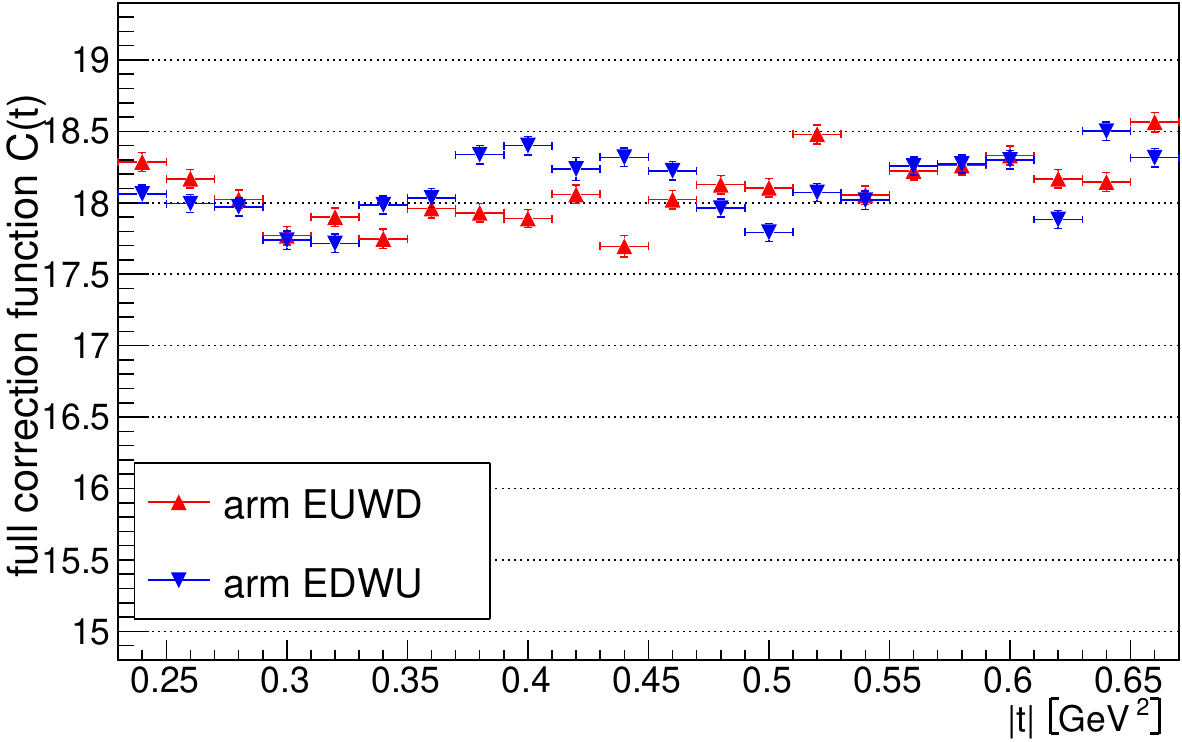}
 \caption{The full correction function $C(t)$ applied to the data
 is shown separately for each detector arm. The horizontal error bars indicate the bin size.}
 \label{fig:Figure4-510GeV}
\end{figure}

The full correction function used to correct the number of reconstructed 4PT elastic events $N_\mathrm{rec}(t)$ is calculated as:
\begin{equation} \label{eq:FullCorFun}
 C(t) = \frac{C_\mathrm{MC}(t) \cdot C_\mathrm{T}(t)}{ \varepsilon_\mathrm{trig}(t)}, 
\end{equation}
where $C_\mathrm{MC}(t) = 1/\varepsilon_\mathrm{MC}(t)$.
Consequently, the differential distribution $(dN/dt)^\mathrm{DATA}$ obtained from data is corrected using a ``bin-by-bin'' method applying the above correction factors:
%
\begin{equation}
 \bigg( \frac{dN}{dt} \bigg)^\mathrm{DATA}_\mathrm{cor} = C(t)\cdot \bigg( \frac{dN}{dt} \bigg)^\mathrm{DATA}_\mathrm{rec}.
 \label{eq:Correction}
\end{equation}
\noindent The values of $C(t)$ for each arm are shown in Fig.~\ref{fig:Figure4-510GeV}. 
One can see that although there is some variation, the factors $\mathrm{C}(t)$ are relatively uniform and are within the range $17.6 < \mathrm{C}(t) < 18.6$ in the $t$ interval of the measurement.
The small modulations observed are due to known individual Si detector plane response behaviors. 

\subsection{MC Weighting Function Correction}
\label{sec:reweighting}
Since the MC is generated with a flat distribution in $t$ while the data has an exponential dependence on $t$, a reweighting of the $dN/dt$ distributions is necessary. A reweighting function based on the FMO model~\cite{Martynov:2018sga}
is used. The systematic uncertainty due to the use of that model is obtained by multiplying the model weighting function by $e^{\pm t}$ and reweighting the $dN/dt$ distributions event by event. The factor $e^{\pm t}$ corresponds to the uncertainty on the slope $B(t)$. The resulting differences in the differential cross section $d\sigma/dt$ are the estimated uncertainties on the differential cross section due to the use of the reweighting function. They are a fraction of the statistical uncertainty and are listed in Table~\ref{tab:dsigmadt-4pt-merg}.

The two major contributions to the systematic uncertainty are those due to the $B$ slope uncertainty used to reweigh the MC sample (MC correction) and to the efficiency correction.
\section{Beam Tilt}
Since the elastic scattering is reconstructed in the RP reference system, additional corrections are needed because of a possible non-zero initial colliding-beam angle or beam tilt in that reference frame.
Such a beam tilt affects the $t$ scale of the measurement.
Note that the offset due to the $(x,y)$ position of the beam at the IP, being a parallel shift, does not change the reconstructed scattering angles $\theta_x, \theta_y$, which are the result of fitting a straight line to the 4PT events.

The beam-tilt angle results in offsets $\tau_{x}$ and $\tau_{y}$ of the reconstructed $\theta_{x}$ and $\theta_{y}$ angles and consequently leads to an offset $\Delta t$ in the calculated $t$-value. In lowest order, where terms proportional to $\tau_{x}^{2}$ and $\tau_{y}^{2}$ are neglected, it is given by:
\begin{equation}\label{eq:DeltatTilt}
 \Delta t \simeq 2 p^{2} (\theta_{x} \tau_{x} + \theta_{y} \tau_{y}).
\end{equation}
In the absence of beam tilt, the ratio of the differential cross sections in two arms  $R(t)=(d\sigma/dt)_\mathrm{EUWD}/(d\sigma/dt)_\mathrm{EDWU}$ is expected to be one. 
The tilt angle $\tau_{x}$ is estimated by forcing the mean value $\langle\theta_{x}\rangle$ of the reconstructed $dN/d\theta_{x}$ to be zero as expected from an elastic event topology, resulting in $\tau_x = 75 \pm 2 \, \mu\mathrm{rad}$, which is confirmed by the MC simulation.

In order to find the $\tau_\mathrm{y}$ we use $\ln{R(|t|)}$, which is a linear function of the difference between the $B(|t|)$ slopes of the two arms. Hence, linear fits to $\ln{R(|t|)}$ are performed in the range  $0.23 \le -t \le 0.67$~GeV$^2$ by iterating the values of $\tau_y$ while $\tau_x$ is set to zero.
When that ratio becomes flat the residual slope is $0.00\pm0.05$. The corresponding systematic uncertainty due to $\tau_y$ is obtained by finding the $\tau_y$ for which the slope of the fitted line is one standard deviation from zero, namely $\pm 0.05$. This procedure yields $\tau_y \approx 20 \pm 5 \, \mu\mathrm{rad}$.
 These values of $(\tau_{x}, \tau_y)$ are then added to the reconstructed values of $(\theta_x,\theta_y)$. 

After the above $(\tau_x,\tau_y)$ corrections, the cross sections $d\sigma/dt$ 
measured separately for the two arms EUWD and EDWU, agree not only in shape but also in normalization to within 2\% level, which is consistent with statistical fluctuations.
The uncertainty on the beam tilt, to which the $\tau_x$ systematic uncertainty contribution is negligible, is propagated to the systematic uncertainty on the cross section measurements and is shown in Table~\ref{tab:dsigmadt-4pt-merg}. It is found to be very small compared to the statistical error.
\section{Background}
\label{sec:Background}
\begin{figure}[!b]
\centering
 \includegraphics[width=1.\columnwidth]{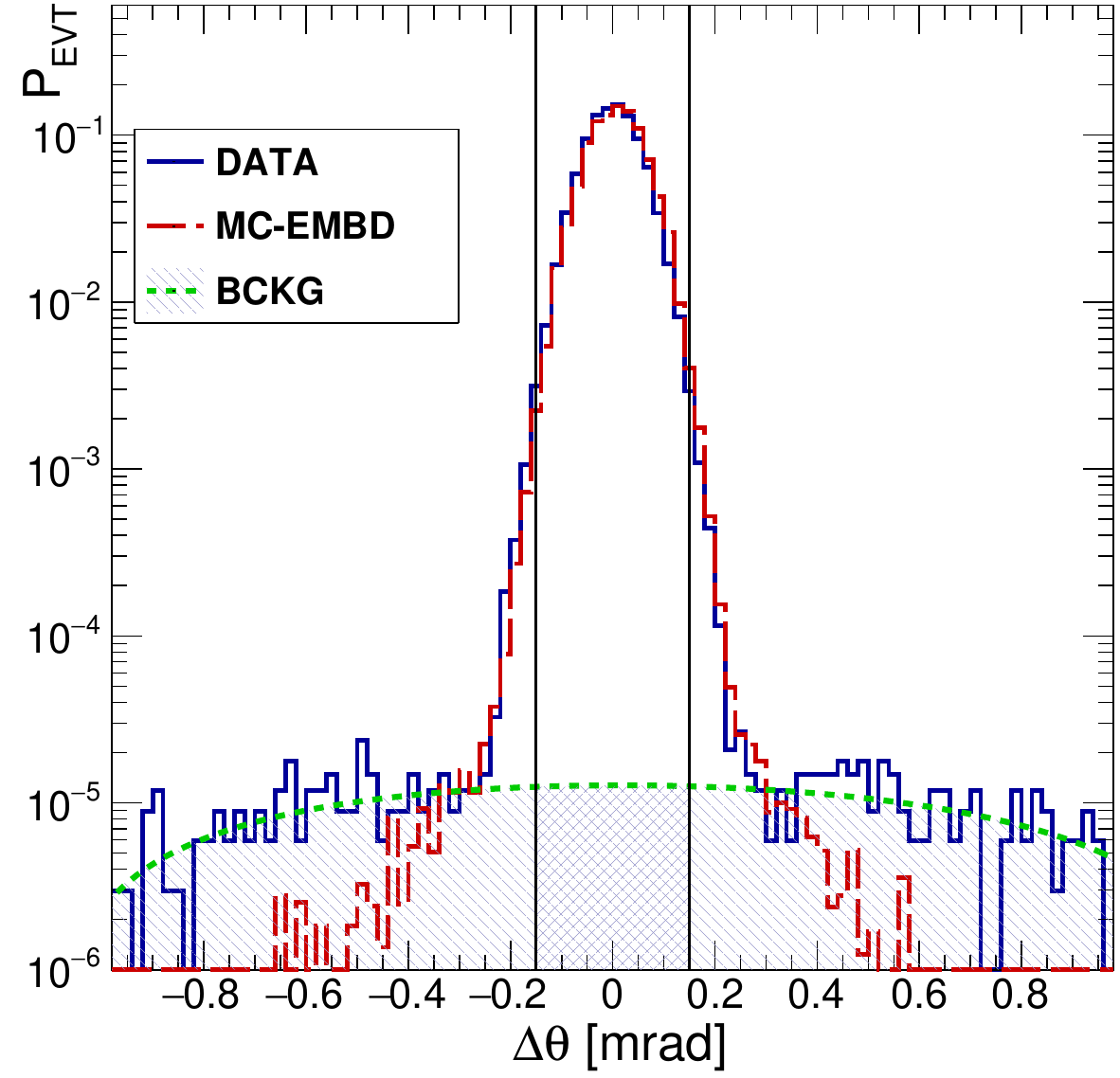}
 \caption{Collinearity, $\Delta\theta=\theta_\mathrm{W}-\theta_\mathrm{E}$, for both GEANT4 based simulation embedded in Zero Bias events (MC-EMBD) and data (DATA) are shown. The vertical axis (P$_\mathrm{EVT}$) is the probability per event.
 Both samples are displayed after the fiducial volume (GEO) cut.
 Estimated background (BCKG), hatched area, and background remaining after collinearity cut, cross-hatched area, are also shown. For the latter, the fraction compared to the signal within the collinearity cut is $0.020 \pm 0.002\%$.}
 \label{fig:Figure5-510GeV}
 \end{figure}

The background estimate is shown in Fig.~\ref{fig:Figure5-510GeV}, where the colli\-nearity distributions for reconstructed data and reconstructed MC samples are compared after the GEO cut. 
In addition to beam halo coincidences and secondary interactions in front of the RPs, the data contains possible physics background sources such as central diffraction and coincidence of two protons from a single diffraction dissociation. 
The collinearity distribution obtained from the GEANT4-based simulation embedded in Zero Bias events is shown as the red-dashed histogram in Fig.~\ref{fig:Figure5-510GeV}, 
which includes background from protons interacting with the material in front of the RPs, such as the beam pipe, magnet structure, the RF shield inside the DX-D0 chamber, etc. 
Since the collinearity distributions for MC and for data are normalized to unity, the vertical axis (P$_\mathrm{EVT}$) in Fig.~\ref{fig:Figure5-510GeV} is the probability per event.

For the background estimate, a second-order polynomial is fitted to collinearity distributions of the data with the exclusion of the $\pm 5\sigma$ central region.
The resulting polynomial is then extrapolated to the central region ($\pm 3\sigma$) of the collinearity distribution, which is the region where the elastic events are used to obtain the cross sections. The background level, compared to the signal within that central phase space is found to be $(0.020 \pm 0.002)\%$, which is negligible and therefore not corrected. The background analysis is repeated in four $t$ regions. The background slightly increases with $|t|$, but is still negligible at the level of 0.1\% at $|t| \approx 0.6$~GeV$^2$.
\section{Results}\label{sec:Results}
\begin{figure}[!b]
\centering
\includegraphics[width=1.\columnwidth]{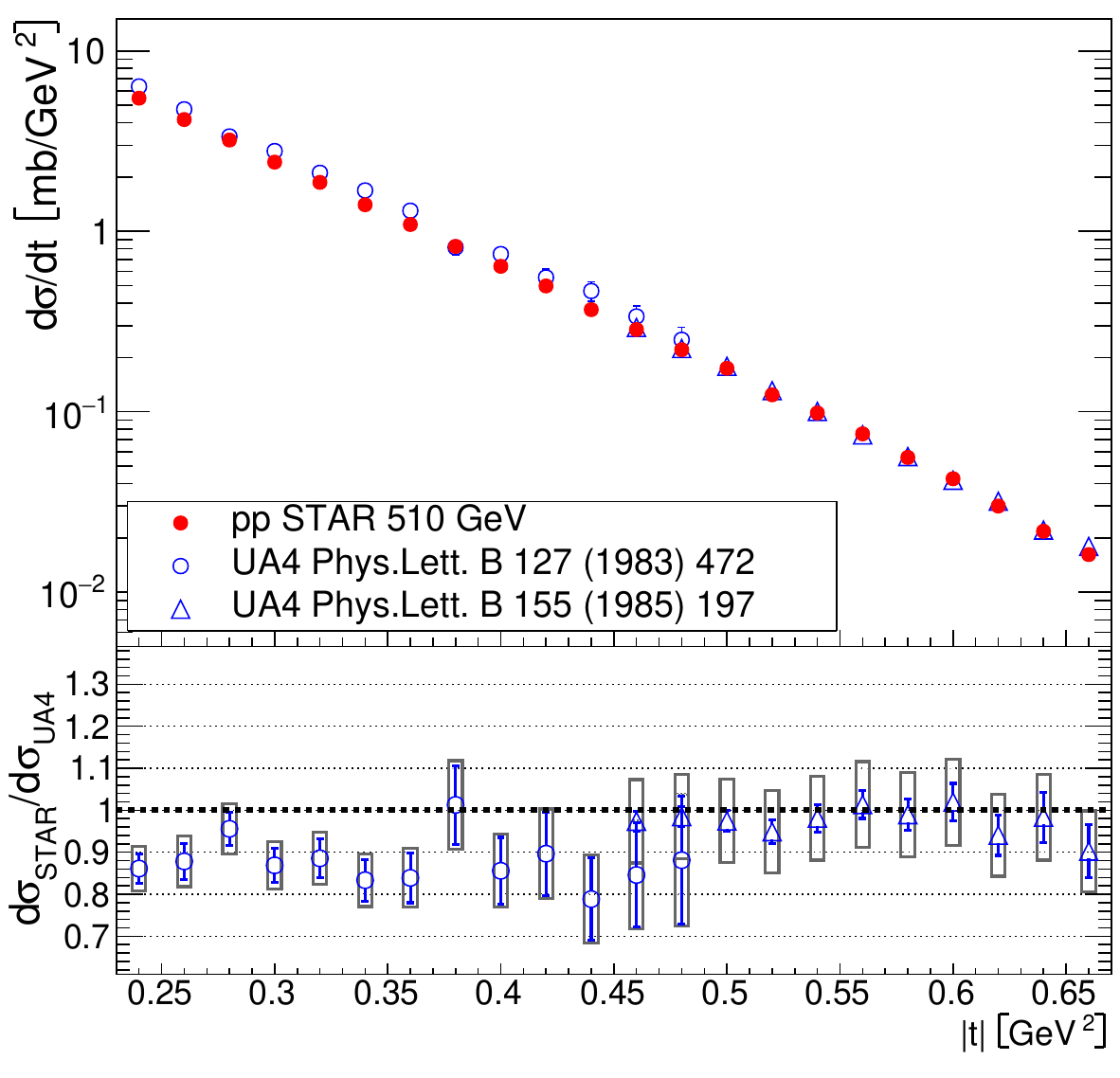}
 \caption{Top panel: The $pp$ elastic differential cross section $d\sigma/dt$ from this experiment compared with the results of the UA4 experiment. Bottom panel: the ratio of the two measurements (STAR/UA4). Uncertainties are statistical only and are smaller than the symbols for the STAR data. 
 The rectangles in the bottom panel show the total systematic uncertainty (STAR + UA4) on the ratio. The vertical scale uncertainty of 2.5\% on the STAR data points, and of 5\% or 10\% (see text) for UA4 data points are not shown here.}
 \label{fig:Figure6-510GeV}
 \end{figure}

\begin{figure}[!b]
\centering
\includegraphics[width=1.\columnwidth]{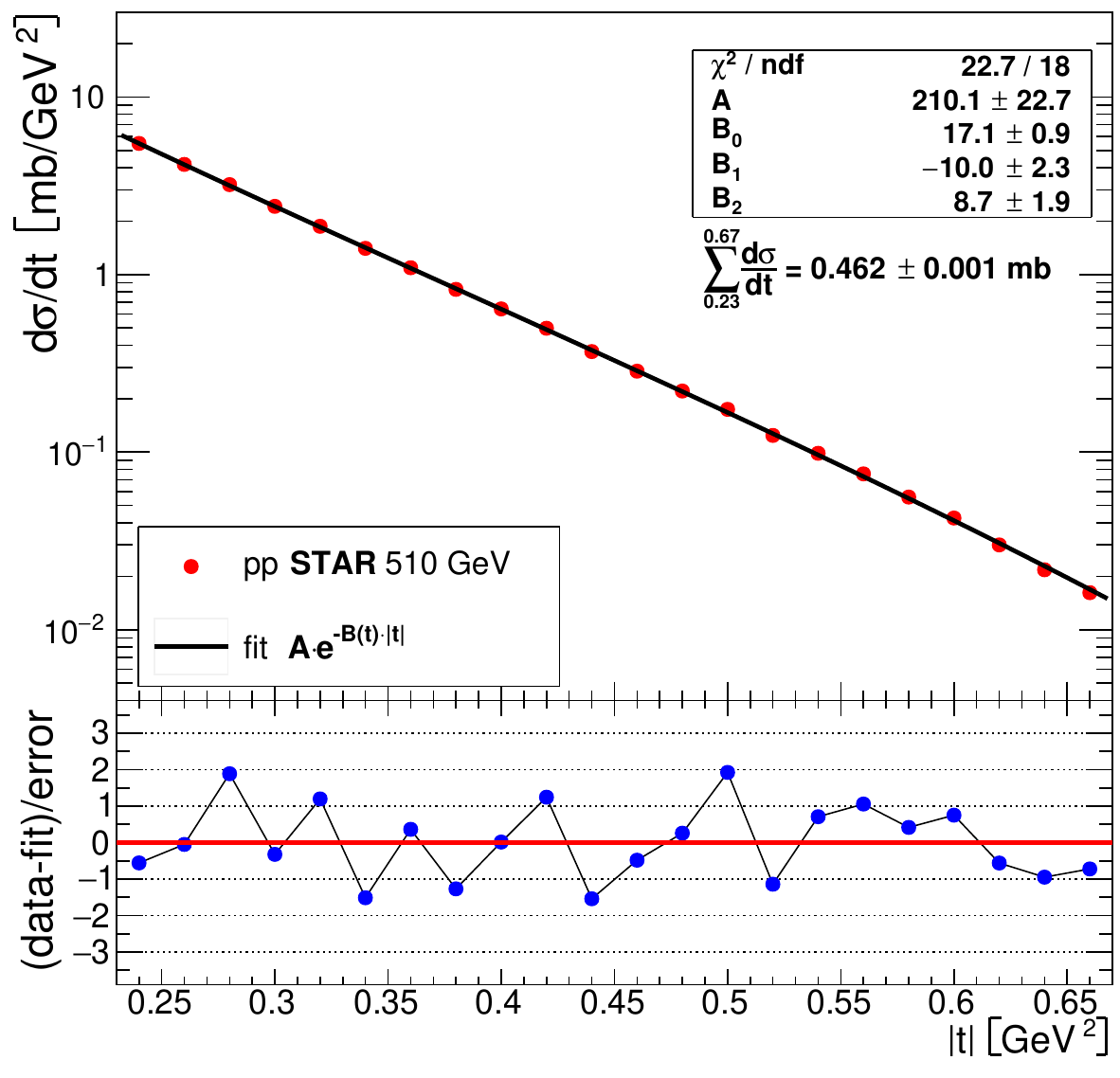}
 \caption{Top panel: The $pp$ elastic differential cross section $d\sigma/dt$ fitted with an exponential $A e^{-B(t)|t|}$ with $B(t)$ as in Eq.~\ref{eq:Boft}. Bottom panel: Residuals (Data - Fit)/Error. Uncertainties on the data points are smaller than the symbol size.}
\label{fig:Figure7-510GeV}
\end{figure}

The final result for $d\sigma/dt$ in the $t$ range of this measurement, $0.23 \leq |t| \leq 0.67$~GeV$^2$, is obtained as a weighted average of $d\sigma/dt$ obtained from each of the two arms of the experiment, where the weights are calculated using statistical errors of the individual points. In Fig.~\ref{fig:Figure6-510GeV} we plot our results and compare them with those from the UA4 experiment for $p\bar p$ at $\sqrt{s} = 540$~GeV. We find a very good agreement between the two measurements as shown in the bottom panel of Fig.~\ref{fig:Figure6-510GeV}, where the ratio is close to one within the experimental uncertainties. 
The systematic shift one observes in the ratio to the UA4 data could be due to a difference in UA4 luminosity normalization, since the associated systematic is reported to be different, which is 5\% for UA4 \cite{UA4:1983mlb} 
for lower $|t|$ values, and 10\% \cite{UA4:1985oqn} 
for larger $|t|$ values.
The statistical and systematic uncertainties on the STAR data points are smaller than the plotted data points, which with their associated uncertainties are presented in Table~\ref{tab:dsigmadt-4pt-merg} and in the HEPData database~\cite{DurhamDB}.
The uncertainties on the ratio are dominated by the uncertainties of the UA4 data.\\
\indent
In the $t$ range of the present measurement, the differential cross section is described by an exponential function $d\sigma_\mathrm{el}/dt = A \cdot \exp{[-B(t)|t|]}$, where $A$ is a normalization constant and $B(t)$ is well approximated by a second order polynomial:
\begin{equation}\label{eq:Boft}
B(t)=  B_0 +B_1\cdot |t| +B_2\cdot |t|^2.
\end{equation}
The $t$ region $|t| < 0.23$~GeV$^2$ is excluded from the analysis due to the significant background contribution from beam halo protons and uncertainty related to detector edge effects.
We find that the fit with the $B(t) = \mathrm{const}$ has a small probability ($\approx 0.001$) and that the quadratic dependence $B(t)$ has a much higher probability of $0.202$. Consequently, we present our result using an exponential fit with $B(t)$ as a second-order polynomial to the measured $d\sigma_\mathrm{el}/dt$. The fit results are shown in Fig.~\ref{fig:Figure7-510GeV} and Table~\ref{tab:FinSysErr}. 

This is the first measurement below the LHC energies for which a non-constant behavior $B(t)$ is observed. The result is in a higher $|t|$ range than the one reported by the TOTEM and ATLAS collaborations \cite{TOTEM:2015oop,ATLAS:2022mgx} at the LHC.

Our result can be compared to that of the UA4 experiment, which found a constant $B$-slope of $13.7 \pm 0.3$~GeV$^{-2}$ in a similar $|t|$-range ($0.21 \leq -t \leq 0.50$~GeV$^2$) to that of this experiment. The UA4 experiment published results from 7,000 events, whereas our sample amounts to 0.35~M events. With the precision here, we are able to determine a non-constant exponential slope in the measured $t$ range.

The integrated elastic scattering cross section, $\sigma^\mathrm{fid}_\mathrm{el}$, within the STAR $t$ acceptance of $0.23 \leq |t| \leq 0.67~\mbox{GeV}^2$ is $\sigma^\mathrm{fid}_\mathrm{el} = 462.1 \pm 0.9 (\mathrm{stat.}) \pm 1.1 (\mathrm {syst.}) \pm 11.6 (\mathrm {scale})$~$\mu\mathrm{b}$.

As described earlier, the estimated background contribution due to particle interactions with the material in front of the RPs and diffractive physics processes within the geometrical acceptance used for this analysis is negligible.
Table~\ref{tab:FinSysErr} contains our results and uncertainty estimates on the exponential fit parameters listed in the left column: the normalization constant~$A$, the slope parameter $B(t)$ and the elastic cross section within STAR's $t$ range $\sigma^\mathrm{fid}_\mathrm{el}$. The systematic uncertainty on the fitted parameters $A$, $B_0$, $B_1$, and $B_2$ is obtained as half of the difference between the fit parameters in the two arms. 
The second to last column of Table~\ref{tab:FinSysErr} lists the total uncertainty, which is calculated by adding the individual uncertainties in quadrature.
For the cross section measurements, the largest systematic uncertainty is the scale uncertainty due to the luminosity determination, which is 2.2\%. The total scale uncertainty, which includes trigger efficiency uncertainty of 1.2\% is 2.5\%.
This scale uncertainty does not affect the value of the slope parameters $B(t)$, but introduces a corresponding systematic uncertainty to the cross sections as listed in Table~\ref{tab:FinSysErr}.
%
\begin{table}[!t]
\centering 
\caption{The measured $pp$ elastic differential cross sections $d\sigma/dt$, with results from the detector arms averaged for the final data sample, as shown in Fig.~\ref{fig:Figure7-510GeV}. The cross section and the center of each $t$ interval, $\boldsymbol{t_\mathrm{cent}}$, are given in [$\mu$b/GeV$^2$] and [GeV$^2$], respectively. The statistical, systematic, and full uncertainties are also listed. The scale uncertainty of 2.5\% is not included in the full error.\\[1pt]} 
\label{tab:dsigmadt-4pt-merg}
\begin{tabular}{c|r|c|c|c} 
\hline 
 $t_\mathrm{cent} $ & d$\sigma$/dt & err. stat. & err. sys. & err. full \\ 
\hline 
 0.24 & 5472.8 & $\pm  23.5$ & $\pm  11.3$& $\pm  26.1$ \\ 
 0.26 & 4167.4 & $\pm  19.6$ & $\pm   8.9$& $\pm  21.5$ \\ 
 0.28 & 3206.8 & $\pm  16.6$ & $\pm   5.9$& $\pm  17.6$ \\ 
 0.30 & 2419.9 & $\pm  13.9$ & $\pm   6.0$& $\pm  15.1$ \\ 
 0.32 & 1868.5 & $\pm  12.0$ & $\pm   4.8$& $\pm  12.9$ \\ 
 0.34 & 1404.4 & $\pm  10.2$ & $\pm   4.1$& $\pm  11.0$ \\ 
 0.36 & 1091.6 & $\pm   8.9$ & $\pm   2.9$& $\pm   9.4$ \\ 
 0.38 &  824.9 & $\pm   7.7$ & $\pm   1.8$& $\pm   7.9$ \\ 
 0.40 &  640.2 & $\pm   6.7$ & $\pm   1.4$& $\pm   6.8$ \\ 
 0.42 &  498.0 & $\pm   5.9$ & $\pm   2.1$& $\pm   6.3$ \\ 
 0.44 &  368.2 & $\pm   5.0$ & $\pm   1.4$& $\pm   5.2$ \\ 
 0.46 &  285.5 & $\pm   4.4$ & $\pm   0.7$& $\pm   4.4$ \\ 
 0.48 &  220.7 & $\pm   3.9$ & $\pm   0.6$& $\pm   3.9$ \\ 
 0.50 &  174.1 & $\pm   3.4$ & $\pm   0.5$& $\pm   3.4$ \\ 
 0.52 &  124.1 & $\pm   2.9$ & $\pm   1.5$& $\pm   3.2$ \\ 
 0.54 &   98.4 & $\pm   2.6$ & $\pm   0.6$& $\pm   2.6$ \\ 
 0.56 &   75.4 & $\pm   2.3$ & $\pm   0.5$& $\pm   2.3$ \\ 
 0.58 &   55.8 & $\pm   1.9$ & $\pm   0.5$& $\pm   2.0$ \\ 
 0.60 &   42.5 & $\pm   1.7$ & $\pm   0.5$& $\pm   1.8$ \\ 
 0.62 &   30.0 & $\pm   1.4$ & $\pm   0.5$& $\pm   1.5$ \\ 
 0.64 &   21.7 & $\pm   1.2$ & $\pm   0.4$& $\pm   1.3$ \\ 
 0.66 &   16.2 & $\pm   1.1$ & $\pm   0.4$& $\pm   1.1$ \\ 
\hline 
\end{tabular} 
\end{table} 
\begin{table}[ht] 
\centering 
\caption{Results of the exponential function fit (Eq.~\ref{eq:Boft}) to the elastic differential cross section data as well as the integrated fiducial cross section are listed. Also listed are the corresponding values of the statistical and systematic uncertainties. The scale (luminosity and trigger efficiency) uncertainty of 2.5\% applicable to the fit parameter \textbf{A} and fiducial cross section $\boldsymbol{\sigma^\mathrm{fid}_\mathrm{el}}$ is not included in the full error.\\[1pt]} 
{\renewcommand{\arraystretch}{1.2}
\begin{tabular}{c|c|r|c|c|c} 
\hline
name  & units & Value  & err. stat. & err. sys. & full  \\
\hline
$\boldsymbol{A}$  & mb/GeV$^{2}$ &  210.1 & $\pm 22.7$ & $ \pm 11.6$ & $\pm 24.1$   \\
$\boldsymbol{B_0}$ & GeV$^{-2}$  &   17.1 & $\pm 0.9$ & $\pm 0.3$ & $\pm 0.9$   \\
$\boldsymbol{B_1}$ & GeV$^{-4}$  &  $-10.0$ & $\pm 2.3$ & $\pm 0.7$ & $\pm 2.4$   \\
$\boldsymbol{B_2}$ & GeV$^{-6}$  &    8.7 & $\pm 1.9$ & $\pm 0.6$ & $\pm 2.0$   \\
$\boldsymbol{\sigma^\mathrm{fid}_\mathrm{el}}$ & $\mu$b & 462.1 & $\pm    0.9$ & $\pm 1.1$& $\pm 1.4$  \\[2pt]
\hline 
\end{tabular}}
\label{tab:FinSysErr} 
\end{table} 
\begin{figure}[!t]
\centering
\includegraphics[width=1.\columnwidth]{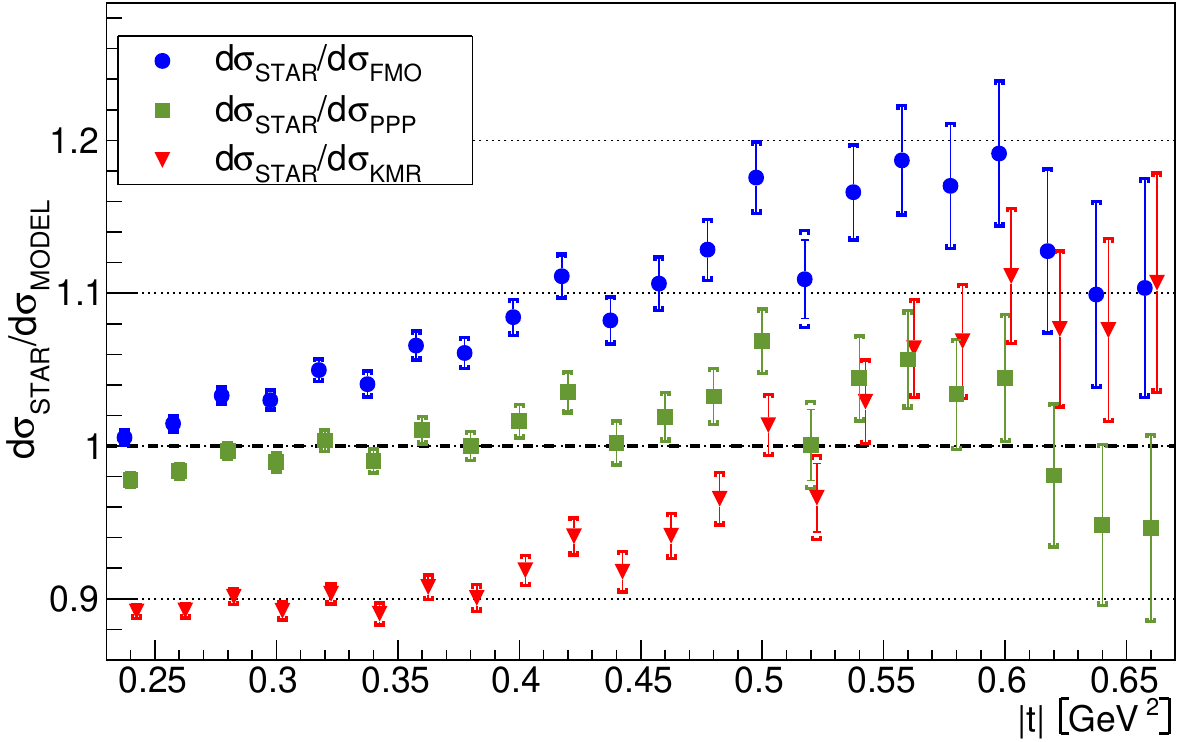}
\caption{Comparison of the STAR $pp$ differential cross section results 
with FMO~\cite{Martynov:2018sga}, KMR~\cite{Khoze:2018kna} and PPP~\cite{Petrov:2002nt} 
models. The ratio of the STAR measured cross section to the model predictions is shown as a function of $t$. The square brackets, where seen, indicate the systematic uncertainty of the STAR data points; otherwise, they are smaller than the symbols themselves.}
\label{fig:Figure8-510GeV}
\end{figure}

In Fig.~\ref{fig:Figure8-510GeV} we compare our $\mathrm{d}\sigma/\mathrm{d}t$ result with three model predictions. The first model (FMO) has a maximum Odderon~\cite{Lukaszuk:1973nt} 
amplitude as described in~\cite{Martynov:2018sga}, 
the second is a two-channel eikonal model (KMR)~\cite{Khoze:2018kna} 
and the third utilizes a three-component Pomeron and an Odderon (PPP)~\cite{Petrov:2002nt}.
We find our result in good agreement with those models, although the agreement is generally better for $|t| < 0.40\; \mathrm{GeV}^2$ than for $|t|$ above that value. 

\begin{figure}[!t]
\centering
\includegraphics[width=1.\columnwidth]{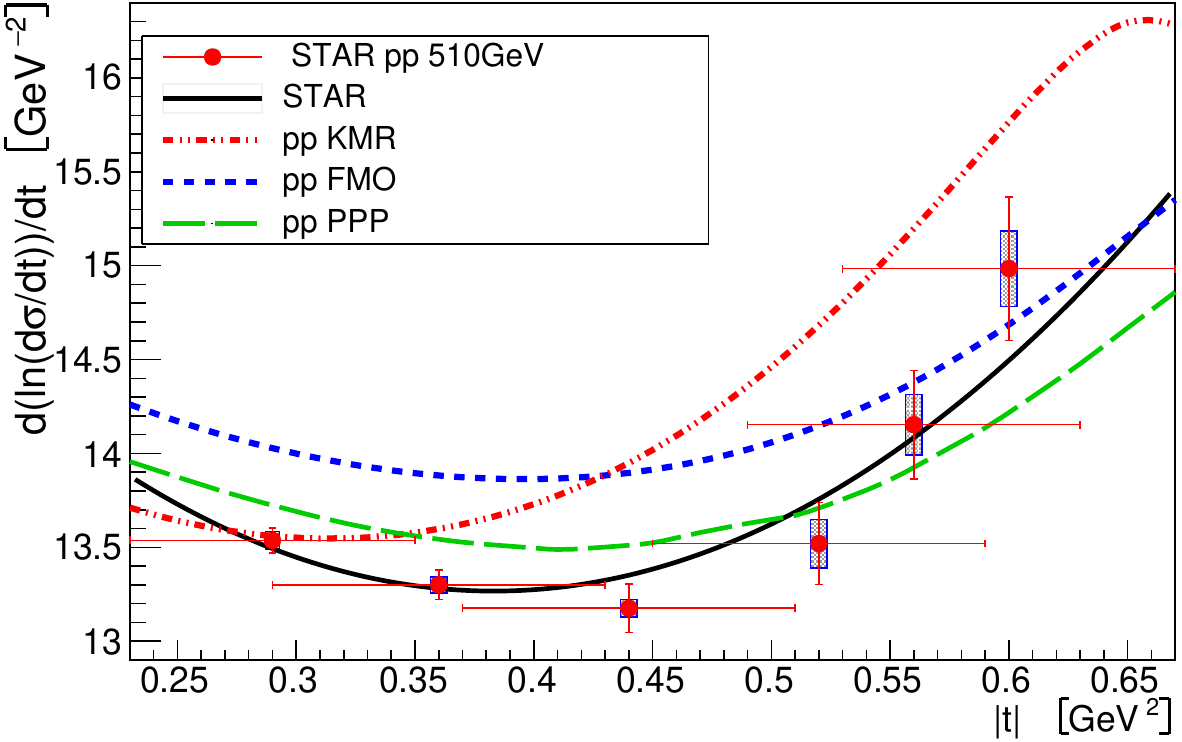}
\caption{Comparison of the STAR $pp$ result in six $t$ sub-intervals with three models: FMO~\cite{Martynov:2018sga}, KMR~\cite{Khoze:2018kna} and PPP~\cite{Petrov:2002nt}.
The vertical axis is $d(\ln{(d\sigma/dt)})/dt$, which is a local slope $B$, if one assumes only a constant term in the exponential.
The black line is fit to the full data set as described in Eq.~\ref{eq:Boft}. The horizontal size of the error bars indicates the $t$ range where $B=\mathrm{const.}$ was fitted. The vertical size of the shaded rectangles indicates the systematic uncertainty of the STAR data points.}
\label{fig:Figure9a-510GeV}
\end{figure}

In order to characterize the shape, we fit a $B=\mbox{const.}$ slope in six sub-intervals of our $t$ range as shown in Fig.~\ref{fig:Figure9a-510GeV}. The vertical axis is a derivative of the logarithm of the differential cross section $d(\ln{(d\sigma/dt)})/dt$, which is a local slope $B$, if one assumes only a constant term in the exponential. There is a good qualitative agreement with the three models shown; there is a minimum in $B(t)$ at $-t \approx 0.40\: \mathrm{GeV}^2$.
\section{Summary}\label{sec:Summary}
We present the STAR experiment's measurement of the elastic differential cross sections for $pp$ scattering at $\sqrt{s} = 510$~GeV as a function of $t$ in the range $ 0.23 \leq |t| \leq 0.67$~GeV$^2$. 
This is the only measurement of the proton-proton elastic cross section in this $t$ range for collision energies above the ISR and below the LHC colliders.
In our $t$ range, the elastic differential cross section is well described by the exponential function $e^{-B(t)|t|}$, where $B(t)$ is a second-order polynomial whose parameters are shown in Fig.~\ref{fig:Figure7-510GeV} and Table~\ref{tab:FinSysErr}.
 This is the first measurement below LHC energies for which a non-constant behavior $B(t)$ is observed. This result is in a higher $|t|$ range than that reported by the TOTEM and ATLAS collaborations at the LHC.
 The UA4 experiment at the $S\!\ppbar S$ collider, at a comparable $\sqrt s$ and $t$ range of this measurement, found a constant $B$-slope, with statistics of 7000 events, compared to 0.35 M of this measurement. The better precision of this experiment allowed to identify the non-constant exponential slope in the measured $t$ range. 

We also present the elastic cross section integrated within the STAR fiducial $t$ range to be $\sigma^\mathrm{fid}_\mathrm{el} = 462.1 \pm 0.9 (\mathrm{stat.}) \pm 1.1 (\mathrm {syst.}) \pm 11.6  (\mathrm {scale})$~$\mu\mathrm{b}$.
We compare $d\sigma_{el}/dt$, in the measured $t$ range, with the results obtained in $\ppbar$ collisions at $\sqrt{s} = 540$~GeV and find that they are in good agreement.
The $\mbox{d}\sigma/\mbox{d}t$ and the shape of $B(t)$, also obtained using six sub-intervals in our $t$ range, are in good agreement with the phenomenological models.
\section*{Acknowledgments}
We thank Prof. E. Martynov, National Academy of Sciences of Ukraine. We thank the RHIC Operations Group and RCF at BNL, the NERSC Center at LBNL, and the Open Science Grid consortium for providing resources and support.  This work was supported in part by the Office of Nuclear Physics within the U.S. DOE Office of Science, the U.S. National Science Foundation, National Natural Science Foundation of China, Chinese Academy of Science, the Ministry of Science and Technology of China and the Chinese Ministry of Education, the Higher Education Sprout Project by Ministry of Education at NCKU, the National Research Foundation of Korea, Czech Science Foundation and Ministry of Education, Youth and Sports of the Czech Republic, Hungarian National Research, Development and Innovation Office, New National Excellency Programme of the Hungarian Ministry of Human Capacities, Department of Atomic Energy and Department of Science and Technology of the Government of India, the National Science Centre and WUT ID-UB of Poland, the Ministry of Science, Education and Sports of the Republic of Croatia, German Bundesministerium f\"ur Bildung, Wissenschaft, Forschung and Technologie (BMBF), Helmholtz Association, Ministry of Education, Culture, Sports, Science, and Technology (MEXT), Japan Society for the Promotion of Science (JSPS) and Agencia Nacional de Investigaci\'on y Desarrollo (ANID) of Chile.

%


\bibliography{ElasticPP510GeVPLBFinal}

\end{document}